\documentclass[]{rAMF2e}
\usepackage{listings}
\usepackage[center]{caption}
\usepackage{centernot}
\usepackage{url}
\usepackage{hyperref}
\usepackage{booktabs}
\usepackage[update,prepend]{epstopdf}
\begin{document}
	\doi{}
	\issn{}  \issnp{}
	\jvol{00} \jnum{00} \jyear{2020} %\jmonth{January--March}
	\def\jobtag{}
	\publisher{Unpublished}
	\jname{}
	
	\markboth{Fabien {Le Floc'h}}{Draft}
	
	\title{More Robust Pricing of European Options Based on Fourier Cosine Series Expansions}
	\author{Fabien {Le Floc'h}}
	\affil{}
	\date{\today}
	\received{v1.1 released June 2020}
	\thanks{Feedback from Matthias Thul, who also naturally implemented the alternative truncation range independently, is gratefully acknowledged.}

\maketitle
\begin{abstract}
	We present an alternative formula to price European options through cosine series expansions, under models with a known characteristic function such as the Heston stochastic volatility model. It is more robust across strikes and as fast as the original COS method.
\end{abstract}

\section{Introduction}
 \citet{fang2008novel} describe a novel approach to the pricing of European options under models with a known characteristic function, based on Fourier cosine series expansions, referred to as the COS method hereafter. The method is very fast but its accuracy is not always reliable for far out-of-the-money options with the proposed truncation range.
 
 We illustrate this issue, explain the root of the error, and derive an alternative pricing formula that stays accurate for out-of-the-money options while staying as fast, even in the case of pricing multiple options of same maturity and different strike prices. 
 
  Confusion around the truncation range is not new, a Wilmott forum post from 2015 \citep{wilmott2015forumcos} suggests the use of an alternative range, and VBA code from 2010 \citep{wilmott2010forumcos} contains the same alternative truncation range setup, although it is not enabled.

 While we investigate only the pricing of European options, which is particularly useful for the calibration of stochastic volatility models, the COS method has also been applied to the pricing of Bermudan and Barrier options \citep{fang2009pricing}, of Asian options \citep{zhang2013efficient} and of two-assets options \citep{ruijter2012two}.
 
 The improvement proposed here is also applicable to the Shannon wavelet SWIFT method from \citet{ortiz2016highly}.

\section{Quick overview of the COS method}
We consider an asset $F$ with a known (normalized) characteristic function
\begin{equation}
\phi(x) = \mathbb{E}\left[e^{i x \ln \frac{F(T,T)}{F(0,T)}}\right]\,.
\end{equation}
$F(0,T)$ is typically the forward price to maturity $T$ of an underlying asset $S$. For example, for an equity with spot price $S$, dividend rate $q$ and interest rate $r$, we have $F(0,T)=S(0)e^{(r-q)T}$. The price of a Put option with the COS method is

\begin{align}
P(K,T) &= B(T) \left[\frac{1}{2}\Re\left(\phi(0)\right)V_0^{\textsf{Put}}+ \sum_{k=1}^{N-1} \Re\left(\phi\left(\frac{k\pi}{b-a}\right)V_k^{\textsf{Put}} e^{ik\pi\frac{-x-a}{b-a}}\right) \right]
\end{align}
with $V_0^{\textsf{Put}} =\frac{2K}{b-a}\left(e^a-1-a\right)$ and for $k \geq 1$
\begin{align}
V_k^{\textsf{Put}} &=\frac{2K}{b-a} \left[ \frac{1}{1+\eta_k^2}\left( e^a + \eta_k \sin\left(\eta_k a\right) - \cos\left(\eta_k a\right)  \right) - \frac{1}{\eta_k} \sin\left(\eta_k a\right) \right] 
\end{align}
where $B(T)$ is the discount factor to maturity, $x = \ln \frac{K}{F(0,T)}$ and $\eta_k=\frac{k\pi}{b-a}$.

The truncation range $[a, b]$ is  chosen according to the first two cumulants $c_1$ and $c_2$ of the model considered  using the rule $a = c_1 - L \sqrt{|c_2|}$, $b = c_1 + L \sqrt{|c_2|}$ and $L$ is a truncation level. The Call option price is obtained through the Put-Call parity relationship.

\section{The problem}
\begin{figure}[htbp]
	\caption{\label{fig:cos_truncation}Error of in-the-money put option prices of maturity 2 days with Heston parameters $\kappa=1.0, \theta=0.1, \sigma=1.0, \rho=-0.9, v_0=0.1, F=1.0$ and different truncation levels $L$.}
	\begin{center}
		\includegraphics[width=\textwidth]{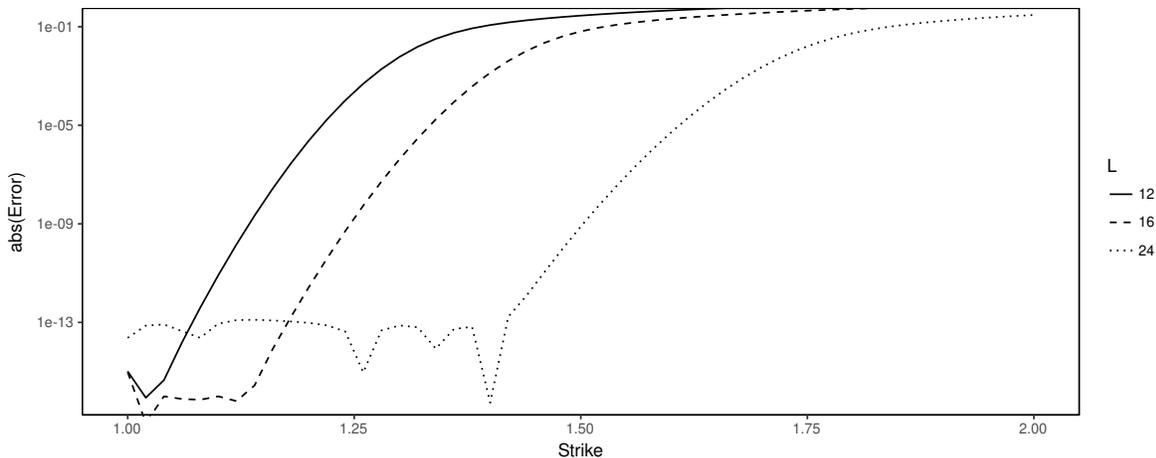}
	\end{center}
\end{figure}

In the COS method, it is particularly important to always compute the Put option price and to rely on the Put Call price parity to keep a high accuracy in general because the absolute value of the cosine coefficients of the Call option increase exponentially with the time to maturity while those of the Put option are constant. We however found out that very in-the-money put option could still be severely mis-priced, depending on the truncation parameter $L$. This is especially visible for very short maturities, even with the recommended value $L=12$. Figure \ref{fig:cos_truncation} shows that the absolute error in the Call option price increases significantly with its strike under the Heston stochastic volatility model. The reference price is obtained by the optimal $\alpha$ method of \citet{lord2007optimal}. Note that when the strike $K$ is beyond the truncation, that is when $\ln \frac{K}{F} \geq b$, the pricing formula is not really applicable anymore and the discounted intrinsic value $B(T)|K-F|^{+}$ should be used instead. Table \ref{tbl:cos_strike_limit} gives the limit strike for different truncation levels $L$.

One way to mitigate the error is to use the discounted intrinsic value before the truncation ends, for example when $x \geq \frac{b}{2}$ or $x \leq  \frac{a}{2}$. But this means it will not be possible to solve the implied volatility is $x$ is too large or too small. In practice, this happens for short maturities and some sets of Heston parameters, even if $L$ is relatively large. We found that including the fourth cumulant numerically did not improve the results. Ideally one would like a larger $L$ for very short maturities and a smaller $L$ for long maturities. 
\begin{table}[h]
	\begin{center}
		\caption{\label{tbl:cos_strike_limit}Strike limits for the parameters of Figure \ref{fig:cos_truncation}}
		\begin{tabular}{c c c c}
			\toprule
			$L$ & $b$ & Strike at $b$ & Strike at $b/2$ \\ 
			\midrule
			12 & 0.2810 & 1.32 & 1.15\\
			16 & 0.3747 & 1.45 & 1.21\\
			24 & 0.5622 & 1.75 & 1.32\\
			\bottomrule
		\end{tabular}
	\end{center}
\end{table}

Why does this happen? The root of this inaccuracy can be found in how the cosine coefficients $V_k^{\textsf{Put}}$ are computed. The coefficients correspond to the cosine transform of the Put payoff:
\begin{align*}
V_k^{\textsf{Put}}&= \frac{2}{b-a} \int_{a}^b K\left|1 - e^y\right|^+ \cos\left(k\pi\frac{y-a}{b-a}\right)dy\\
&=\frac{2}{b-a} \int_{a}^0 K\left(1 - e^y\right) \cos\left(k\pi\frac{y-a}{b-a}\right)dy\\
&=\frac{2}{b-a} K\left(-\chi_k(a,0)+\psi_k(a,0)\right)
\end{align*}
with $y=\ln \frac{S_T}{K}$ and the functions $\chi$ and $\psi$ defined in \citep[p. 6]{fang2008novel} equations (22) and (23).

In particular, the cosine coefficients for the Put option are computed relatively to the strike price $K$ but the truncation range is relative to the spot price.

\section{An alternative pricing formula}
An alternative is to compute the cosine coefficients of the Put option relative to the forward price $F$:
\begin{align}
V_k^{\textsf{Put}}&= \frac{2}{b-a} \int_{a}^b F\left|\frac{K}{F} - e^y\right|^+ \cos\left(k\pi\frac{y-a}{b-a}\right)dy\\
&=\frac{2F}{b-a} \int_{a}^z \left(e^z - e^y\right) \cos\left(k\pi\frac{y-a}{b-a}\right)dy\\
&=\frac{2F}{b-a}\left(-\chi_k(a,z)+e^z\psi_k(a,z)\right)\\
&=\frac{2}{b-a} \left(-F\chi_k(a,z)+K\psi_k(a,z)\right)
\end{align}
where $z=\ln \frac{K}{F}$ and $y=\ln \frac{S_T}{F}$.

We have 
\begin{align}
V_0^{\textsf{Put}}(z)&=2F\frac{e^a-e^z+e^z(z-a)}{b-a}\,,\\
V_k^{\textsf{Put}}(z)&=  \frac{2F}{(b-a)\left(1+\eta_k^2\right)}\left[e^a - \cos\left(\eta_k(z-a)\right)e^z - \eta_k\sin\left(\eta_k(z-a)\right)e^z  \right]\nonumber\\
&\quad+ \frac{
2F}{(b-a)\eta_k} \sin\left(\eta_k(z-a)\right)e^z \quad \textmd{ for } k=1,...,N-1\label{eqn:cos_f_vk}
\end{align}
with $\eta_k = \frac{k\pi}{b-a}$.
%TODO F or S0

The Put option price is then obtained by the usual formula, but with $x=0$.
\begin{equation}
P(F,K,T) = B(T) \left[\frac{1}{2}\Re\left(\phi(0)\right)V_0^{\textsf{Put}}(z)+ \sum_{k=1}^{N-1} \Re\left(\phi\left(\frac{k\pi}{b-a}\right) e^{-i k \pi\frac{a}{b-a}}\right)V_k^{\textsf{Put}}(z) \right]\,.\label{eqn:cos_f_price}
\end{equation}
Contrary to the original COS method, the $V_k^{\textsf{Put}}$ coefficients now depend on the strike and thus need to be recomputed for each strike. In the evaluation of $\psi$ and $\chi$, the costliest operation is to compute the cos and sin functions. This needs to be done for each $k$. But now the term $\Re\left(\phi\left(\frac{k\pi}{b-a}\right) e^{-i k \pi\frac{a}{b-a}}\right)$ is fully independent of the strike and can be pre-computed, for each maturity. This saves one cos and one sin function evaluation per $k$. The total cost is thus the same as the original COS method.

It can be verified that this alternative formula is equivalent to shifting the truncation range from $[a, b]$ to $[a-\ln \frac{K}{F},b - \ln\frac{K}{F}]$ in the original COS method, but using equations (\ref{eqn:cos_f_vk})  and (\ref{eqn:cos_f_price}) is much more efficient to compute option prices for a range of strikes.

When the strike $K$ is such that $z < a$, the Put option value should be set directly to 0. Similarly When $z > b$, the Put option value should be set to its intrinsic value.
\section{Numerical Example}
We consider the same short maturity options as in Figure \ref{fig:cos_truncation}, and we compute the absolute error in the price of a Call option with a truncation level $L=12$ of the classic and the improved method for $N=256$, varying the strike.
\begin{figure}[h]
	\caption{\label{fig:cosk}Absolute error of in-the-money put option prices of maturity 2 days using $\kappa=1.0, \theta=0.1, \sigma=1.0, \rho=-0.9, v_0=0.1, F=1.0$ for the truncation level $L=12$.}
	\begin{center}
		\includegraphics[width=.9\textwidth]{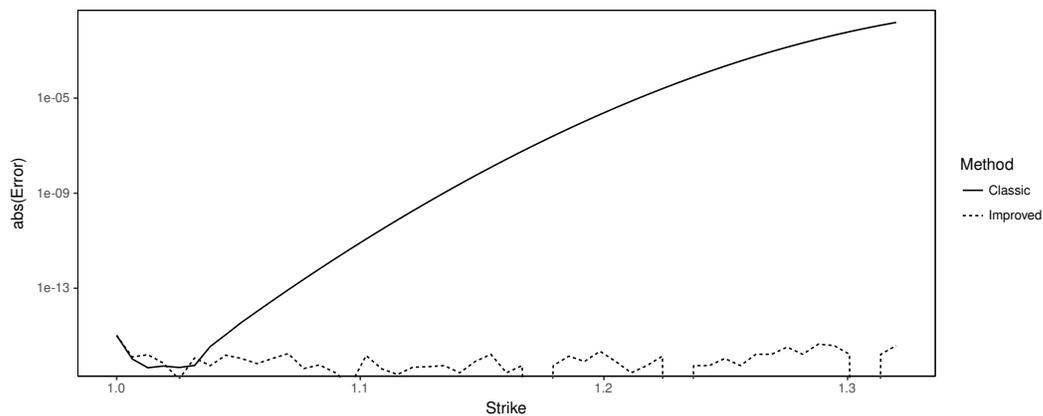}
	\end{center}
\end{figure}
 We stop at strike $K=1.32$ since then $\ln \frac{K}{F} >b$. Figure \ref{fig:cosk} shows that the error of the improved method stays below $10^{-15}$, close to machine epsilon while the error of the classic method can be as high as $1.5\cdot 10^{-2}$.

For longer maturities, and well behaved Heston parameters, it turns out that the new formula has a constant error over the range of strikes, while the classic formula has a lower error for $K<F$ (Puts) and much larger for $K>F$ (Calls). In appendix \ref{sec:error_estimate}, we show that the error of the COS method $e(z)$ is composed of two terms:
\begin{align}
 e(z)&=\int_{\mathbb{R}\setminus[a,b]}\left(v(y,T)-\hat{v}(y,T) \right) f(y|x) dy+ \sum_{k=N}^{\infty}\Re\left[\phi\left(\frac{k\pi}{b-a}\right)e^{ik\pi\frac{-a}{b-a}}\right]V_k
\end{align}
where $\hat{v}$ is the cosine expansion of the payoff $v$ and $f$ is the probability density.
The first term is the payoff approximation error beyond the boundaries and the second term is the series truncation error. When $N$ is sufficiently large, the first error will dominate.

\begin{figure}[h]
	\caption{Error in the Put payoff approximation $v(z,T)-\hat{v}(z,T)$ for $a =-3.45125, b= 3.025$ and $F=2016$, which corresponds to the interval for the Black model with volatility $\sigma=55\%$, maturity $T=1.0$ and truncation level $L=6$.}
	\begin{center}
			\subfigure[][Low strikes $z < a$.\label{fig:cos_black_payoffa}]{\includegraphics[width=.90\textwidth]{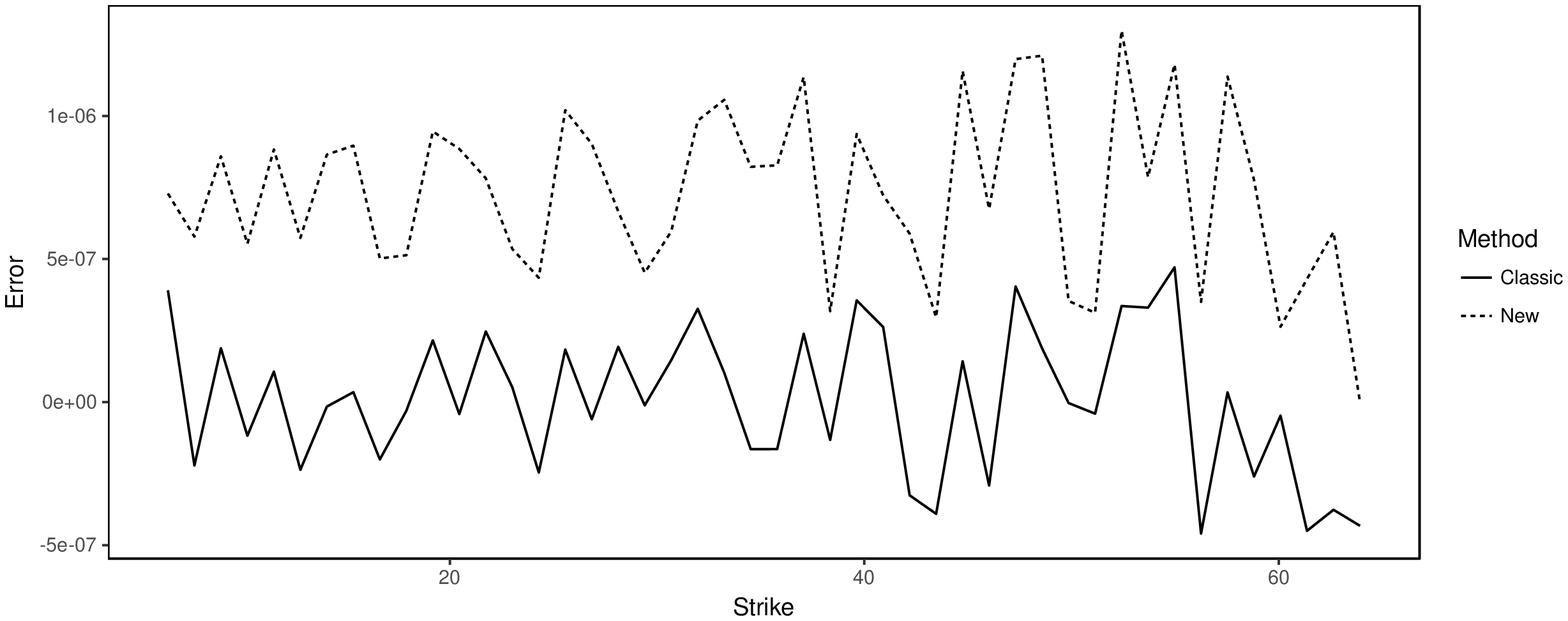}}
		\subfigure[][Classic, high strikes $z>b$.\label{fig:cos_black_payoffb_classic}]{
			\includegraphics[width=0.45\textwidth]{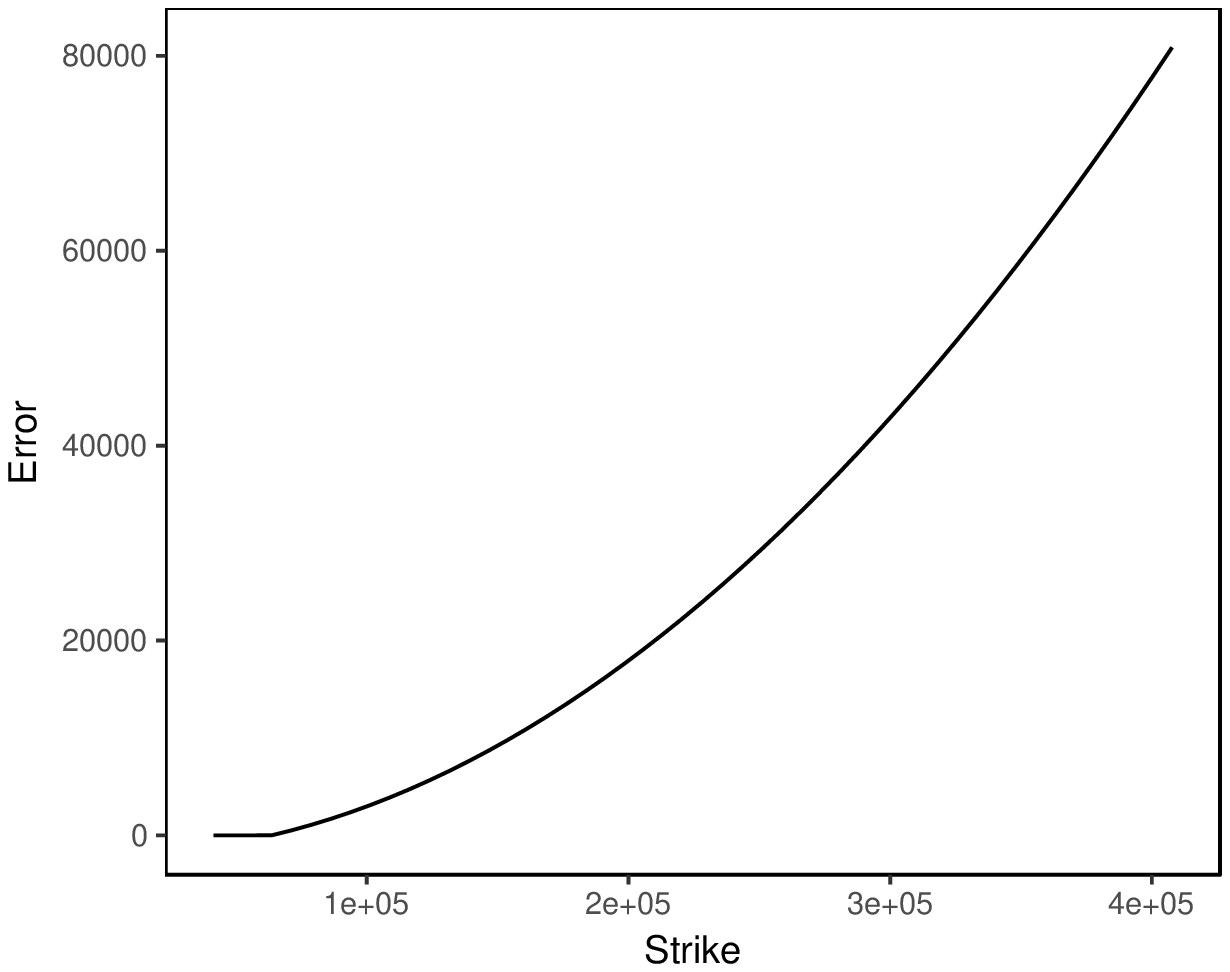}}
		\subfigure[][New, high strikes $z>b$.\label{fig:cos_black_payoffb_new}]{
			\includegraphics[width=0.45\textwidth]{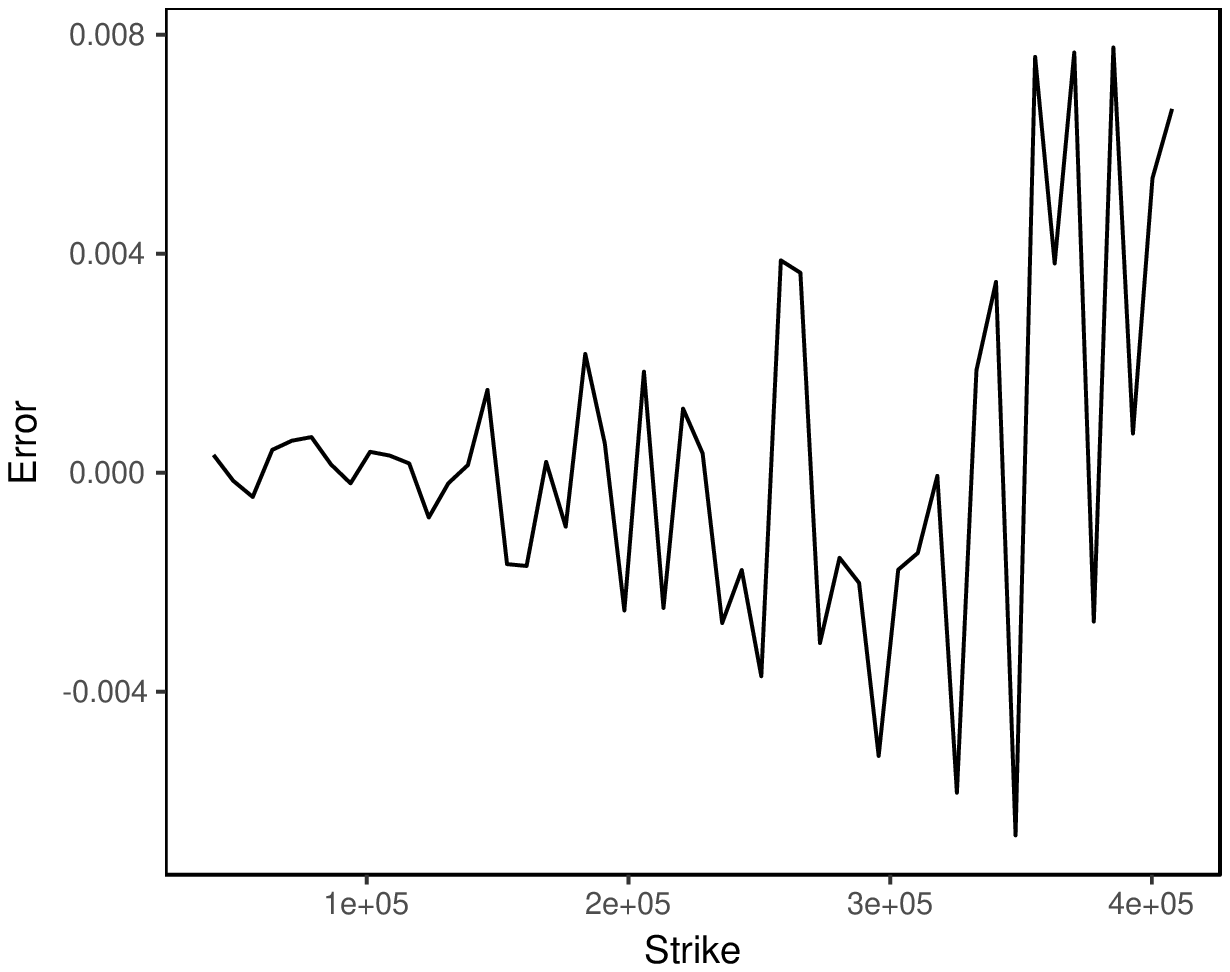}}	
	\end{center}
\end{figure}

In Figure \ref{fig:cos_black_payoffa}, we plot the payoff error $v-\hat{v}$ by strike both the classic and the new method for $K< F e^{a}$.
We plot the payoff error separately for large strikes $K>F^{b}$ in Figures \ref{fig:cos_black_payoffb_classic} and \ref{fig:cos_black_payoffb_new} since the scale is very different between the two approaches.

The payoff approximation of the classic method is slightly better for low strikes, especially since it oscillates around zero, which will results some cancellation. But its amplitude for large strikes is alarming and explains why the overall error of the classic method degrades for $K>F$. In contrast, the payoff error of the new method oscillates around zero for high strikes.

\subsection{Challenging Heston parameters}
We consider an option of maturity $T=1$ and strike $K=0.25$ on an asset following the Heston stochastic volatility model with parameters $v_0 = 0.0225, \kappa= 0.1, \theta=0.01, \sigma=2.0, \rho= 0.5, F=1$. The option is therefore very out of the money. The reference price is given by the Lord and Kahl optimal $\alpha$ method \citep{lord2007optimal} and we compute prices with the classic and the improved Cos method with a truncation level $L=12$ and a large $N=16384$ so that series truncation is not an issue. On this example, we include the fourth cumulant ($c_4=0.05827$) in the calculation of the truncation range, as it plays an important role ($c_1=-0.01095$, $c_2=0.01808$).  Note that our choice of truncation level is relatively pessimistic since \citet{fang2008novel} recommend $L=8$.

The results presented in Table \ref{tbl:heston_medium} show that the Cos method with the new coefficients presented in this paper significantly reduces\footnote{In a previous version, we made an error in the calculation of the truncation range, which led to a much more accurate price for lower strikes on this example. This is corrected here.} the maximum error over a range of strikes. Indeed, the absolute error stays relatively constant then. In comparison, the original formula leads to higher accuracy for low strikes and much worse error for high strikes, as the truncation range is effectively shifted.

\begin{table}[h]
	\caption{out-of-the-money 1Y option of notional 1,000,000 with Heston parameters $v_0 = 0.0225, \kappa= 0.1, \theta=0.01, \sigma=2.0, \rho= 0.5$ under various numerical methods. The option is a put for strikes lower than 100\%, and a call otherwise (priced by put-call parity).\label{tbl:heston_medium}}
	\centering{
		%	\begin{small}
		\begin{tabular}{r c c r r r}
			\toprule
			Strike & Method & Tolerance & Points &  Price & Error \\\midrule
			25\% & \texttt{modlob} Lord-Kahl & 1e-8 & 7907 & 119.38532 & 0.00000\\
&Cos (Classic) &L=12& 16384& 119.38531 & 0.00002 \\
&Cos (Improved) &L=12& 16384& 119.38418 &0.00115 \\
			50\% & \texttt{modlob} Lord-Kahl & 1e-8 & 4497 & 834.40773& 0.00000\\
			 &Cos (Classic) &L=12& 16384& 834.40756 & 0.00017 \\
			 &Cos (Improved) &L=12& 16384&834.40657  &0.00116 \\
			100\% & \texttt{modlob} Lord-Kahl & 1e-8 & 467 & 20511.93508& 0.00000\\
&Cos (Classic) &L=12& 16384& 20511.93388 & 0.00120 \\
&Cos (Improved) &L=12& 16384&20511.93388  &0.00120 \\			 
			200\% & \texttt{modlob} Lord-Kahl & 1e-8 & 3787 & 6563.82888& 0.00000\\
&Cos (Classic) &L=12& 16384& 6563.82102 & 0.00786 \\
&Cos (Improved) &L=12& 16384&6563.82773  &0.00115 \\			 
			400\% & \texttt{modlob} Lord-Kahl & 1e-8 & 6537 & 3951.92085& 0.00000\\
&Cos (Classic) &L=12& 16384& 3951.86703 & 0.05382 \\
&Cos (Improved) &L=12& 16384&3951.91908  &0.00177 \\			 
			 \bottomrule
		\end{tabular}
	}
	%\end{small}
\end{table}

\section{Conclusion}
For the same computational cost as the original COS method, a more uniform accuracy  across strikes is obtained with the alternate COS formula presented in this paper. This is particularly visible for deep out-of-the-money options.
%\section{Conclusion}
\bibliographystyle{rAMF}
\bibliography{cos_method_improved}

\appendix
\section{Error estimate}\label{sec:error_estimate}
Let $v(x,t)$ be the undiscounted option price at time $t$ and $f(y|x)$ the probability density of being at $y$ starting from $x$. At maturity $T$ the payoff is $v(x,T)$. For a European (non-path dependent) option we can price the option by integrating over the density:
\begin{equation}
v(x,t)=\int_{-\infty}^{+\infty}v(y,T)  f(y|x) dy
\end{equation}
In order to evaluate the integral, we will truncate it to an interval $[a,b]$:
\begin{equation}
v(x,t)=\int_{\mathbb{R}\setminus[a,b]}v(y,T)  f(y|x) dy + \int_{a}^{b}v(y,T)  f(y|x) dy \\
\end{equation}
We now use the cosine expansion on the interval $[a,b]$: $f(y|x) = \sum_{k=0}^{\infty}{'} A_k(x)\cos\left(k\pi\frac{y-a}{b-a}\right)$
with 
\begin{align}
A_k(x) = \frac{2}{b-a} \int_{a}^b f(y|x) \cos\left(k\pi\frac{y-a}{b-a}\right) dy
\end{align}
This leads to 
\begin{align}
v(x,t)&= \int_{\mathbb{R}\setminus[a,b]}v(y,T)  f(y|x) dy + \int_{a}^{b} v(y,T) \sum_{k=0}^{\infty}{'} A_k \cos\left(k\pi\frac{y-a}{b-a}\right) dy
\end{align}
Numerically, the sum stops at a finite $N \in \mathbb{N}$. We thus split the sum in two parts:
\begin{align}
v(x,t)&=\int_{\mathbb{R}\setminus[a,b]}v(y,T)  f(y|x) dy +\int_{a}^{b} v(y,T) \sum_{k=N}^{\infty} A_k \cos\left(k\pi\frac{y-a}{b-a}\right) dy\nonumber\\
&\quad+ \int_{a}^{b} v(y,T) \sum_{k=0}^{N-1}{'} A_k \cos\left(k\pi\frac{y-a}{b-a}\right) dy
\end{align}
The characteristic function $\phi$ corresponding to the density $f$ is $\phi(x)=\int_{-\infty}^{+\infty} e^{iux} f(u)du$. We thus have the identity
\begin{equation} \Re\left[\phi\left(\frac{k\pi}{b-a}\right)e^{ik\pi\frac{-a}{b-a}}\right] =\int_{-\infty}^{+\infty} \cos\left(k\pi\frac{u-a}{b-a}\right) f(u)du\end{equation}

We will use it in the definition of $A_k$ to obtain
\begin{align}
	A_k &= \frac{2}{b-a}\int_{\mathbb{R}} f(y|x) \cos\left(k\pi\frac{y-a}{b-a}\right) dy -  \frac{2}{b-a}\int_{\mathbb{R}\setminus[a,b]}  f(y|x) \cos\left(k\pi\frac{y-a}{b-a}\right) dy\\
	&=\frac{2}{b-a}\Re\left[\phi\left(\frac{k\pi}{b-a}\right)e^{ik\pi\frac{-a}{b-a}}\right] - \frac{2}{b-a}\int_{\mathbb{R}\setminus[a,b]}  f(y|x) \cos\left(k\pi\frac{y-a}{b-a}\right)  dy
\end{align}
We replace $A_k$ in the last integral of $v(x,t)$ to obtain 
\begin{align}
v(x,t)&=\int_{\mathbb{R}\setminus[a,b]}v(y,T)  f(y|x) dy +\int_{a}^{b} v(y,T) \sum_{k=N}^{\infty} A_k \cos\left(k\pi\frac{y-a}{b-a}\right) dy\nonumber\\
&\quad- \sum_{k=0}^{N-1}{'} \frac{2}{b-a} \left[\int_{\mathbb{R}\setminus[a,b]}  f(y|x) \cos\left(k\pi\frac{y-a}{b-a}\right) dy \right] \int_{a}^{b} v(y,T) \cos\left(k\pi\frac{y-a}{b-a}\right) dy\nonumber \\
&\quad+\sum_{k=0}^{N-1}{'}\Re\left[\phi\left(\frac{k\pi}{b-a}\right)e^{ik\pi\frac{-a}{b-a}}\right] \frac{2}{b-a}\int_{a}^{b} v(y,T)   \cos\left(k\pi\frac{y-a}{b-a}\right) dy
\end{align}

The cosine expansion of the payoff is $\hat{v}(y,T) = \sum_{k=0}^{\infty} V_k \cos\left(k\pi\frac{y-a}{b-a}\right)$ with 
\begin{equation}V_k =  \frac{2}{b-a} \int_{a}^b v(y,T) \cos\left(k\pi\frac{y-a}{b-a}\right) dy\end{equation}
For  $y \in [a,b]$ we have $v(y) = \hat{v}(y)$.
This leads to
\begin{align}
v(x,t)&=\int_{\mathbb{R}\setminus[a,b]}v(y,T)  f(y|x) dy +\int_{a}^{b} v(y,T) \sum_{k=N}^{\infty} A_k \cos\left(k\pi\frac{y-a}{b-a}\right) dy\nonumber\\
&\quad- \sum_{k=0}^{N-1}{'}  \left[\int_{\mathbb{R}\setminus[a,b]}  f(y|x) \cos\left(k\pi\frac{y-a}{b-a}\right) dy \right]V_k\nonumber\\
&\quad+\sum_{k=0}^{N-1}{'}\Re\left[\phi\left(\frac{k\pi}{b-a}\right)e^{ik\pi\frac{-a}{b-a}}\right] V_k
\end{align}
Contrary to what is done in \cite[equation (43)]{fang2008novel} (corrected in \citep{fang2010cos}), we can not expand $v(y,T)$ as a cosine serie in the first term since the expansion matches the original payoff $v(y,T)$ only inside the interval $[a,b]$. Outside, it is periodic, while the original payoff is not. Instead,we will decompose $v$ as $v-\hat{v} + \hat{v}$.

\begin{align}
v(x,t)&=\int_{\mathbb{R}\setminus[a,b]}\left(v(y,T)-\hat{v}(y,T) \right) f(y|x) dy\nonumber\\
&\quad+\int_{\mathbb{R}\setminus[a,b]}\hat{v}(y,T)  f(y|x) dy+\int_{a}^{b} v(y,T) \sum_{k=N}^{\infty} A_k \cos\left(k\pi\frac{y-a}{b-a}\right) dy\nonumber\\
&\quad- \sum_{k=0}^{N-1}{'}  \left[\int_{\mathbb{R}\setminus[a,b]}  f(y|x) \cos\left(k\pi\frac{y-a}{b-a}\right) dy \right]V_k\nonumber\\
&\quad+\sum_{k=0}^{N-1}{'}\Re\left[\phi\left(\frac{k\pi}{b-a}\right)e^{ik\pi\frac{-a}{b-a}}\right] V_k
\end{align}
Now we expand $\hat{v}$ in the second integral, this simplifies with the fourth integral to obtain
\begin{align}
v(x,t)&= \int_{\mathbb{R}\setminus[a,b]}\left(v(y,T)-\hat{v}(y,T) \right) f(y|x) dy\nonumber\\
&\quad+\int_{\mathbb{R}\setminus[a,b]} f(y|x) \sum_{k=N}^{\infty}V_k   \cos\left(k\pi\frac{y-a}{b-a}\right)   dy+\int_{a}^{b} v(y,T) \sum_{k=N}^{\infty} A_k \cos\left(k\pi\frac{y-a}{b-a}\right) dy\nonumber\\
&\quad+\sum_{k=0}^{N-1}{'}\Re\left[\phi\left(\frac{k\pi}{b-a}\right)e^{ik\pi\frac{-a}{b-a}}\right] V_k
\end{align}
We recognize the definition of $V_k$ in the third integral, this means:
\begin{align}
v(x,t)&= \int_{\mathbb{R}\setminus[a,b]}\left(v(y,T)-\hat{v}(y,T) \right) f(y|x) dy\nonumber\\
&\quad+\int_{\mathbb{R}\setminus[a,b]} f(y|x) \sum_{k=N}^{\infty}V_k   \cos\left(k\pi\frac{y-a}{b-a}\right)   dy + \sum_{k=N}^{\infty} A_k \frac{b-a}{2}V_k\nonumber\\
&\quad+\sum_{k=0}^{N-1}{'}\Re\left[\phi\left(\frac{k\pi}{b-a}\right)e^{ik\pi\frac{-a}{b-a}}\right] V_k
\end{align}
We now use the definition of $A_k$ in the third integral to obtain
\begin{align}
v(x,t)&= \int_{\mathbb{R}\setminus[a,b]}\left(v(y,T)-\hat{v}(y,T) \right) f(y|x) dy\nonumber\\
&\quad+\int_{\mathbb{R}\setminus[a,b]} f(y|x) \sum_{k=N}^{\infty}V_k   \cos\left(k\pi\frac{y-a}{b-a}\right)   dy+\sum_{k=N}^{\infty} V_k \int_{a}^b f(y|x) \cos\left(k\pi\frac{y-a}{b-a}\right) dy\nonumber\\
&\quad+\sum_{k=0}^{N-1}{'}\Re\left[\phi\left(\frac{k\pi}{b-a}\right)e^{ik\pi\frac{-a}{b-a}}\right] V_k
\end{align}
We can now combine the second and third integrals together:

\begin{align}
	v(x,t)&= \int_{\mathbb{R}\setminus[a,b]}\left(v(y,T)-\hat{v}(y,T) \right) f(y|x) dy+\sum_{k=N}^{\infty}V_k\int_{\mathbb{R}} f(y|x)  \cos\left(k\pi\frac{y-a}{b-a}\right)   dy\nonumber\\
	&\quad+\sum_{k=0}^{N-1}{'}\Re\left[\phi\left(\frac{k\pi}{b-a}\right)e^{ik\pi\frac{-a}{b-a}}\right] V_k
\end{align}
We recognize the characteristic function identity in the second integral and we obtain
\begin{align}
v(x,t)&=\int_{\mathbb{R}\setminus[a,b]}\left(v(y,T)-\hat{v}(y,T) \right) f(y|x) dy+ \sum_{k=N}^{\infty}\Re\left[\phi\left(\frac{k\pi}{b-a}\right)e^{ik\pi\frac{-a}{b-a}}\right]V_k\nonumber\\
&\quad+\sum_{k=0}^{N-1}{'}\Re\left[\phi\left(\frac{k\pi}{b-a}\right)e^{ik\pi\frac{-a}{b-a}}\right] V_k
\end{align}

\section{First Cumulants}
\subsection{First cumulants for Heston}
The two first cumulants $c_1$ and $c_2$ of $\log\left(\frac{F}{K}\right)$ are used to define the integration boundaries $a$ and $b$ of the COS method.
The cumulant generating function is:
\begin{equation}
g(u) = \log(\phi(-iu))
\end{equation}
We have 
\begin{align}
c_1 &= g'(0)\\
c_2 &= g''(0)
\end{align}
Those can be computed numerically. Analytic formulas are given in \citep{fang2008novel}, unfortunately their formula for $c_2$ is wrong. Here are our own derived formulas from a Taylor expansion:
\begin{equation}\label{eqn:c1h}
c_1 = (1-e^{-\kappa t})\frac{\theta-v_0}{2\kappa}-\frac{1}{2}\theta t
\end{equation}
\begin{equation}\label{eqn:c2h}
\begin{split}
c_2 = \frac{v_0}{4\kappa^3}\{ 4 \kappa^2 \left(1+(\rho\sigma t -1)e^{-\kappa t}\right)   + \kappa \left(4\rho\sigma(e^{-\kappa t}-1)-2\sigma^2 t e^{-\kappa t}\right)+\sigma^2(1-e^{-2\kappa t}) \}\\
+  \frac{\theta}{8\kappa^3} \{ 8 \kappa^3 t - 8 \kappa^2 \left(1+ \rho\sigma t + (\rho\sigma t-1)e^{-\kappa t}\right) + 2\kappa \left( (1+2e^{-\kappa t})\sigma^2 t+8(1-e^{-\kappa t})\rho\sigma \right) \\
+ \sigma^2(e^{-2\kappa t} + 4e^{-\kappa t}-5) \}
\end{split}
\end{equation}

The fourth cumulant, derived by a computer algebra system, reads $c_4 = c_4^A + c_4^B$ with
\begin{align*}
c_4^B =&\frac {2{\sigma}^{2} v_0}{{\kappa}^{7}} \left(  \left(  \left( {T}^{3}{\rho}^{3}\sigma-3\,{T}^{2}{\rho}^{2} \right) {\kappa}^{6}-3/2\,T \left( {T}^{2}{\rho}^{2}{\sigma}^{2}-2\,T\rho\, \left( {\rho}^{2}+2 \right) \sigma+4\,{\rho}^{2}+2 \right) {\kappa}^{5} \right. \right.\\
&\left.\left. + \left( 3/4\,{T}^{3}\rho\,{\sigma}^{3}-6\,{T}^{2} \left( {\rho}^{2}+3/8 \right) {\sigma}^{2}+6\,T\rho\, \left( {\rho}^{2}+2 \right) \sigma-6\,{\rho}^{2} \right) {\kappa}^{4} \right.\right.\\
&\left.\left. -1/8\,\sigma\, \left( {T}^{3}{\sigma}^{3}-24\,{T}^{2}\rho\,{\sigma}^{2}+ \left( 72\,T{\rho}^{2}+18\,T \right) \sigma-48\,{\rho}^{3} \right) {\kappa}^{3} \right.\right.\\
&\left.\left. -3/8\,{\sigma}^{2} \left( {T}^{2}{\sigma}^{2}-7\,T\sigma\,\rho-3 \right) {\kappa}^{2}-3/16\,{\sigma}^{3} \left( T\sigma+10\,\rho \right) \kappa  +3/8\,{\sigma}^{4} \right) {{\rm e}^{-\kappa\,T}} \right.\\
&\left. + \left(  \left( -3/2-3\,{T}^{2}{\rho}^{2}{\sigma}^{2}+6\,T\sigma\,\rho \right) {\kappa}^{4}+3\,\sigma\, \left( {T}^{2}\rho\,{\sigma}^{2}+ \left( -3\,T{\rho}^{2}-3/2\,T \right) \sigma+3\,\rho \right) {\kappa}^{3} \right.\right.\\
&\left.\left. -3/4\,{\sigma}^{2} \left( {T}^{2}{\sigma}^{2}-10\,T\sigma\,\rho+12\,{\rho}^{2}+3 \right) {\kappa}^{2}-{\frac { \left( 9\,T\sigma-30\,\rho \right) {\sigma}^{3}\kappa}{8}}-3/8\,{\sigma}^{4} \right) {{\rm e}^{-2\,\kappa\,T}} \right.\\
&\left.+{\frac {9\,{\sigma}^{2} \left(  \left( T\sigma\,\rho-1 \right) {\kappa}^{2}+ \left( -1/2\,{\sigma}^{2}T+5/3\,\rho\,\sigma \right) \kappa-1/3\,{\sigma}^{2} \right) {{\rm e}^{-3\,\kappa\,T}}}{8}} \right.  \\
&\left. -{\frac {3\,{\sigma}^{4}{{\rm e}^{-4\,\kappa\,T}}}{32}}-6\, \left( \kappa\,\rho\,\sigma-1/4\,{\sigma}^{2}-{\kappa}^{2} \right)  \left(  \left( {\rho}^{2}+1/4 \right) {\kappa}^{2}-5/4\,\kappa\,\rho\,\sigma+{\frac {5\,{\sigma}^{2}}{16}} \right)  \right) 
	\end{align*}
and	
\begin{align*}
c_4^A =& -\frac {2{\sigma}^{2}\theta}{{\kappa}^{7}} \left(  \left(  \left( {T}^{3}{\rho}^{3}\sigma-3\,{T}^{2}{\rho}^{2} \right) {\kappa}^{6}-3/2\,T \left( {T}^{2}{\rho}^{2}{\sigma}^{2}-4\,T\rho\, \left( {\rho}^{2}+1 \right) \sigma+8\,{\rho}^{2}+2 \right) {\kappa}^{5} \right. \right.\\
&\left. \left. + \left( 3/4\,{T}^{3}\rho\,{\sigma}^{3}-21/2\,{T}^{2} \left( {\rho}^{2}+3/14 \right) {\sigma}^{2}+ \left( 18\,T{\rho}^{3}+24\,T\rho \right) \sigma-18\,{\rho}^{2} \right)  {\kappa}^{4} \right. \right.\\
&\left.\left. -1/8\, \left( {T}^{3}{\sigma}^{3}-42\,{T}^{2}\rho\,{\sigma}^{2}+ \left( 240\,T{\rho}^{2}+54\,T \right) \sigma-192\,{\rho}^{3}-192\,\rho \right) \sigma\,{\kappa}^{3}\right.\right.\\
&\left.\left.  -3/4\,{\sigma}^{2} \left( {T}^{2}{\sigma}^{2}-{\frac {35\,T\sigma\,\rho}{2}}+40\,{\rho}^{2}+15/2 \right) {\kappa}^{2}  -{\frac {27\,{\sigma}^{3}\kappa}{16} \left( T\sigma-{\frac {20\,\rho}{3}} \right) }-{\frac {21\,{\sigma}^{4}}{16}} \right) {{\rm e}^{-\kappa\,T}} \right.\\
&\left. + \left(  \left( -3/4-3/2\,{T}^{2}{\rho}^{2}{\sigma}^{2}+3\,T\sigma\,\rho \right) {\kappa}^{4}  +3/2 \sigma\, \left( {T}^{2}\rho\,{\sigma}^{2}+ \left( -4\,T{\rho}^{2}-3/2\,T \right) \sigma+4\,\rho \right) {\kappa}^{3} \right.\right.\\
&\left.\left. -3/8\,{\sigma}^{2} \left( {T}^{2}{\sigma}^{2}-14\,T\sigma\,\rho+20\,{\rho}^{2}+6 \right) {\kappa}^{2}  + \left( -{\frac {15\,T{\sigma}^{4}}{16}}+9/2\,\rho\,{\sigma}^{3} \right) \kappa-{\frac {21\,{\sigma}^{4}}{32}} \right) {{\rm e}^{-2\,\kappa\,T}} \right.\\
&\left. +3/8\,{\sigma}^{2} \left(  \left( T\sigma\,\rho-1 \right) {\kappa}^{2}+ \left( -1/2\,{\sigma}^{2}T+2\,\rho\,\sigma \right) \kappa-1/2\,{\sigma}^{2} \right) {{\rm e}^{-3\,\kappa\,T}}-{\frac {3\,{\sigma}^{4}{{\rm e}^{-4\,\kappa\,T}}}{128}} \right.\\
&\left. + \left( -3/2\,T-6\,T{\rho}^{2} \right) {\kappa}^{5}+ \left(  \left( 6\,T{\rho}^{3}+9\,T\rho \right) \sigma+18\,{\rho}^{2}+{\frac{15}{4}} \right) {\kappa}^{4}\right.\\
&\left. -9\,\sigma\, \left( T \left( {\rho}^{2}+1/4 \right) \sigma+8/3\,{\rho}^{3}+10/3\,\rho \right) {\kappa}^{3} \right.\\
&\left. +{\frac {15\,{\sigma}^{2}{\kappa}^{2}}{4} \left( T\sigma\,\rho+10\,{\rho}^{2}+{\frac{11}{5}} \right) }+ \left( -{\frac {33\,\rho\,{\sigma}^{3}}{2}}-{\frac {15\,T{\sigma}^{4}}{32}} \right) \kappa+{\frac {279\,{\sigma}^{4}}{128}} \right) 
	\end{align*}
Alternatively, Taylor series algorithmic differentiation \citep{neidinger2013efficient} may be used to compute the fourth cumulant as suggested in \citep{thul2016cos}.

\subsection{First cumulants for stochastic volatility with jump (SVJ)}

A Taylor expansion of the SVJ cumulant generating function around 0 leads to:
\begin{align}
c_1 &= c_{1_H} + (\alpha-\bar{k}) \lambda T\\
c_2 &= c_{2_H} + \left( \alpha^2+\delta^2 \right)\lambda T\\
c_4 &= c_{4_H} + \left( \alpha^4+6\delta^2 \alpha^2+3\delta^4 \right)\lambda T
\end{align}

%4alpha2 + -8alphadelta2 +4delta2 
with $\alpha = \log(1+\bar{k})-\frac{1}{2}\delta^2$ and $c_{1_H}, c_{2_H}$ are the Heston cumulants given in Equations (\ref{eqn:c1h}) and (\ref{eqn:c2h}).

In practice, we use in the COS method $\hat{c}_2 = c_2 + \sqrt{|c_4|}$ making the approximation $c_{4_H} = 0$ to take into account the 4th cumulant effect of the jump part only. This is particularly important as jumps can introduce a relatively high fourth cumulant.

\subsection{First cumulants for double Heston}

By linearity, the first cumulants for double Heston are just the sum of the Heston cumulants corresponding to each volatility process.

\end{document}